\documentstyle[preprint,aps,prb]{revtex}
\begin{document}
\draft

\title{Infrared and Raman Evidence for Dimers and Polymers in Rb$_1$C$_{60}$}

\author{Michael C. Martin$^1$, Daniel Koller$^1$, A. Rosenberg$^2$,
C. Kendziora$^2$, and L. Mihaly$^1$}
\address{$^1$Department of Physics, SUNY at Stony Brook, Stony Brook, NY
11794-3800\\
$^2$Code 6650, Naval Research Laboratory, Washington, DC  20375}

\date{05 Aug 1994, revised 26 Oct 1994, Accepted in Phys. Rev. B}
\maketitle

\begin{abstract}
The infrared- and Raman-active vibrational modes of C$_{60}$ were measured
in the various structural states of Rb$_1$C$_{60}$.
According to earlier studies, Rb$_1$C$_{60}$ has an $fcc$ structure at
temperatures above $\sim 100^\circ$C, a linear chain polymer orthorhombic
structure when slowly cooled, and an as yet undetermined structure when
very rapidly cooled (``quenched").
We show that the
spectra obtained in the polymer state are consistent with each C$_{60}$
molecule having bonds to two diametrically opposite neighbors.
In the quenched state, we find evidence for further symmetry
breaking, implying a lower symmetry structure than the polymer state.
The spectroscopic data of the quenched phase are shown
to be consistent with Rb$_2$(C$_{60}$)$_2$, a dimerization of C$_{60}$.
\end{abstract}
\pacs{PACS: 78.30.Jw, 61.46.+w, 78.66.Qn}


A$_1$C$_{60}$ was not discovered \cite{win92} until well after the other
phases of alkali intercalated C$_{60}$ were known, and it was found to
undergo a number of structural and electronic phase
transitions.\cite{win92,poi93,jan93,zhu93,tyc93,rb1,cha94,ste94}
The existence of a phase consisting of linear polymer chains,\cite{ste94}
which is stable in air,\cite{stable,peksci} has
stimulated recent interest in this material and could help in
developing applications for alkali-doped fullerenes.

At high temperatures $(\agt 100^\circ $C), Rb$_1$C$_{60}$ has an $fcc$
rocksalt structure \cite{zhu93} and is electrically
conducting.\cite{poi93,jan93,tyc93,rb1}
When slowly cooled, the specimen undergoes a first-order phase
transition to an orthorhombic phase,\cite{cha94} forming linear chains
of C$_{60}$ molecules along the $a$ direction (the face-diagonal
direction of the $fcc$ phase).\cite{ste94}  These chains
have been proposed to be $[2+2]$ cycloaddition C$_{60}$ polymers.\cite{pek94}
When Rb$_1$C$_{60}$ samples are cooled extremely rapidly from the $fcc$ phase
to below $\sim 0^\circ$C a metastable phase is produced.\cite{jan93,rb1}
This ``quenched" phase is insulating \cite{rb1} and relaxes back to the
stable polymer chain phase, with a strongly temperature dependent
timescale.\cite{jan93,rb1}  Dimers forming from $[2+2]$
cycloaddition or a single bond between two free radical C$_{60}$ molecules
have been suggested as a possible structure.\cite{pekker}
To date, x-ray scattering structural studies of this phase are
still incomplete.\cite{faigel}  Vibrational spectroscopy, however,
can help determine the symmetry of the C$_{60}$ molecules in this
quenched phase by observing which vibrational modes are infrared- (IR)
or Raman-active.

In this communication we show, for the first time, that the IR and Raman
vibrational spectra of the slowly cooled and quenched states are indeed
consistent with polymerized and dimerized C$_{60}$ structures, respectively.
Using this C$_{60}$ dimer picture, we are able to show that many of the
characteristic properties of this phase are understood.

The C$_{60}$ films used for the present study were evaporated and doped in
vacuum-sealed pyrex sample chambers described in detail
previously.\cite{chamb}  Si substrates were utilized for the IR measurements,
and MgO substrates were used for the Raman measurements.  During doping,
IR spectra obtained {\it in situ} were used to monitor the stoichiometry of
all samples by measuring the position of the $F_{1u}$(4) vibrational
mode.\cite{c60dope,rice}  Once Rb$_x$C$_{60}$ samples with the majority phase
$x=1$ were produced, the alkali appendages were removed from the pyrex
chambers, leaving permanently
sealed chambers of essentially known stoichiometry.  These rugged,
portable chambers allowed a number of measurements under a variety of
thermal conditions.  A copper block with a heater resistor and
thermocouple was attached to each chamber in order to regulate and measure
the sample temperature by means of a temperature controller.  Quenching
was accomplished by pouring liquid nitrogen directly on the sample chambers.
Typically, the sample temperature dropped from 125$^\circ$C to
-200$^\circ$C in approximately 30 seconds.

IR spectra were obtained at 2cm$^{-1}$ resolution on a Bomem MB-155 Fourier
transform spectrometer and ratioed to spectra of the empty sample chamber at
the same temperature.  Raman spectroscopy was performed using the 514.5nm
Ar$^+$ laser line and a Dilor XY triple monochromator with a CCD detector
at a resolution of $\sim 5$cm$^{-1}$; care was taken to use low laser
powers and to monitor the $A_g(2)$ mode when the sample was in a
temperature range where photoinduced effects (photopolymerization or
photoexcitation) of C$_{60}$ have been observed \cite{photo}
(the Raman-active $A_g(2)$ mode shifts when C$_{60}$ becomes photopolymerized
or photoexcited).

The IR transmission spectra obtained for all structures are presented in
Figure \ref{IR}.  The $fcc$ data (lowest curve) reveal only the four
$F_{1u}$-derived vibrational modes, as labeled.  The $F_{1u}(4)$ mode shows
small amounts of pure C$_{60}$ and Rb$_6$C$_{60}$ , labeled $x=0$ and $x=6$
respectively, in the mostly Rb$_1$C$_{60}$ sample (the $F_{1u}(4)$
absorption of the $x=6$ compound appears large in Fig. \ref{IR} only
because the mode in the $x=6$ material is much stronger than in other
stoichiometries \cite{c60dope,rice}).  The approximate center positions
(in cm$^{-1}$) of all observed absorptions of the slow-cooled (middle
curve) and quenched (top curve) states are labeled in the figure.
Note that the overall transmission of the quenched phase is very
different from those of the $fcc$ and polymerized phases (the $fcc$ and
polymerized phases are electrically conducting and have low transmissions,
while the quenched phase is insulating and has a high transmission
\cite{rb1}).  For easier comparison, the IR spectra have been scaled such
that the vibrational absorptions due to small amounts of other
stoichiometric phases (labeled $x=0$ and $x=6$ in Fig. \ref{IR}) appear
approximately equal in magnitude.  Since the strengths of these
``impurity" modes should be independent of the state of the predominant
Rb$_1$C$_{60}$ in any given sample, this scaling allows meaningful visual
comparisons of the modes seen in each phase.

We present the Raman spectra for all three states in Figure \ref{raman}.
The bottom panel shows the $fcc$ phase
spectrum with the ten known Raman vibrational modes of C$_{60}$ labeled;
these appear unchanged from the pure C$_{60}$ vibrations (while the sample
is known to be majority $x=1$ phased by IR spectroscopy, this Raman spectrum
is probably dominated by a surface layer of the $x=0$ ``impurity" phase).
The middle panel
displays the spectrum of the polymerized state, with modes that are not
seen in the $fcc$ state labeled.  Inset to this panel is a drawing of
the fullerenes in their polymer structure (the full orthorhombic
Rb$_1$C$_{60}$ structure is shown in Ref. \cite{ste94}).  The top panel
displays the spectrum of the quenched phase, with one of the proposed
fullerene dimers (the $[2+2]$ cycloaddition possibility) shown in the inset.

We begin the discussion with the polymer chain structured state (drawn in
the inset to the middle panel of Fig. \ref{raman}).  A drastic
change in molecular environment occurs when the samples go through the
first-order phase transition from the high temperature rocksalt
structure, where the C$_{60}$ molecules are only bound by Van der Waals
forces, to the polymer state where hexagon-hexagon (6-6) bonds on the
C$_{60}$ cages are broken and bonds between neighboring fullerenes are
formed.  This change can be phrased in group theoretical terms as a symmetry
lowering from the highly symmetric $I_h$ icosohedral C$_{60}$ structure to
the much less symmetric $D_{2h}$ group of the polymer.  Since the center of
inversion site symmetry is preserved, the Raman and IR
measurements remain strictly complementary.  Applying the $D_{2h}$ site
symmetry to the fullerene molecule, one expects all the {\it ungerade}
(antisymmetric) vibrations to become IR-active and similarly all the
{\it gerade} (symmetric) modes to become Raman-active.  Thus, the number
of IR- and Raman-active modes will be greatly increased over fullerene
compounds with no C$_{60}$-C$_{60}$ bonds (such as the $fcc$ phase of
Rb$_1$C$_{60}$).

In agreement with this prediction, the Rb$_1$C$_{60}$ polymer IR spectrum
in Figure \ref{IR} indeed reveals at least 12 absorptions, in addition to
the four $F_{1u}$-derived modes, which must have {\it ungerade} symmetry.
Analogously, the Raman spectrum of this
phase (middle panel of Fig. \ref{raman}) also reveals many newly
Raman-active vibrations which must be {\it gerade} modes.  Since the parity
of these modes is determined, these results can help to sort out the
symmetries of IR- and Raman-forbidden vibrational modes of pure C$_{60}$
seen in previous measurements.\cite{c60modes,neutron,eklund}

In the polymer state, in addition to the appearance of these new modes,
the $F_{1u}(3)$ mode at 1182cm$^{-1}$ splits into two modes at 1183 and
1197cm$^{-1}$, and the $F_{1u}(4)$ mode at 1395cm$^{-1}$ splits into modes
at 1388 and 1407cm$^{-1}$ .  These splittings are expected since
vibrational motion along the polymer chain
direction is perturbed much more than motion perpendicular to the
chains.  The $F_{1u}$ vibrations are three-fold degenerate with motions
along the translational axes (x, y, and z); polymerization breaks this
degeneracy and a split mode is observed.  An alternate explanation for
these splittings has been offered by Pederson and Quong \cite{peddimer}:
a mixing occurs between the z-rows of the nearly degenerate $G_u(6)$ and
$F_{1u}(4)$ states, creating a perturbed IR-active mode, and similarly the
$F_{1u}(3)$ state mixes with the $F_{2u}(4)$ and $H_u(5)$ modes of their
theoretical model.  Either explanation is consistent with both the
existence of the polymer state and the spectroscopic data.

Turning to the IR and Raman spectra of the quenched phase (top curves
of Fig. \ref{IR} and Fig. \ref{raman}, respectively), we note a number
of differences relative to the polymer phase.  First, many more
vibrational absorptions are seen to have become IR-active (at least 25
in addition to the four $F_{1u}$-derived modes, as compared to 12 new modes
in the polymer state).  Similarly, the Raman spectrum of this quenched state
shows several more vibrational modes than are observed in the polymer state.
Second, the $F_{1u}(3)$ and $F_{1u}(4)$ vibrational modes are again observed
to split, but less than in the polymer case:  the splittings are
9cm$^{-1}$ for the $F_{1u}(3)$ line and 6cm$^{-1}$ for
the $F_{1u}(4)$ line in the quenched phase, compared to the polymer state
splittings of 14 and 19cm$^{-1}$ , respectively.  These observations
indicate that a new and different type of symmetry breaking is occurring.

We compare our data to modes expected in the proposed quenched state
structure \cite{pekker} composed of C$_{60}$ dimers, Rb$_2$(C$_{60})_2$.
A simple sketch
of two fullerene molecules in such a dimerized state is shown in the inset
to the top panel of Figure \ref{raman}.  The exact type of bond between
the two C$_{60}$ balls will have to be determined by further structural
measurements.  The sketch in Fig. \ref{raman} shows a $[2+2]$ cycloaddition
bond similar to the polymer state bonds, but a single bond 
or other bond types are equally possible and do not change the present
results.  This dimerized state has even lower site symmetry than the
polymer state:  the center of inversion is no longer at the center of
a C$_{60}$ molecule.  Group theory then predicts that all {\it gerade}
and {\it ungerade} vibrations will be both IR- and Raman-active in the
quenched state.

To test this prediction, above Figure \ref{IR} we have placed vertical lines
at the positions of the Raman-active $A_g$ and $H_g$ modes of pure C$_{60}$
which fall within our measurement range.  Previous measurements have shown
that most vibrational modes of C$_{60}$ soften upon doping with alkali
metals.\cite{mitch,c60dope,kuz,rice}  We indeed observe modes in the IR
spectra of the quenched phase that are not IR-active in the polymer phase
at energies just below the ideal $A_g$ and $H_g$ energies, in agreement
with Raman studies of A$_1$C$_{60}$ \cite{mitch} (although it is unclear
which state of Rb$_1$C$_{60}$ was measured in that study).  Likewise,
$F_{1u}$ modes appear in the Raman spectrum of the quenched phase (top
curve of Fig. \ref{raman} at 527, 566cm$^{-1}$) but not in the
polymer state Raman spectrum.  These observations strongly suggest that
the center of inversion
symmetry is broken in the quenched state, in agreement with the
dimerization picture.  This allows us to assign all modes observed in the
IR spectra of the quenched state that are not also in the polymer state as
being derived from {\it gerade} symmetry C$_{60}$ modes.  Likewise, the
modes in the Raman spectra which are seen in the quenched phase and not in
the polymer phase have {\it ungerade} symmetry.

The dimer picture is in qualitative agreement with the smaller
observed splittings of the $F_{1u}(3)$ and $F_{1u}(4)$ modes.  Since each
fullerene is altered only on one side when dimerized, as opposed to two
sides when
polymerized, the perturbation to the C$_{60}$ eigenmodes will be smaller.
The insulating nature of this quenched state can also be understood to be
due to dimerization.  The LUMO (lowest unoccupied molecular orbital)
in the uncharged dimer is derived from a splitting of the triply
degenerate $t_{1u}$ electronic orbital of pure C$_{60}$ into a singly
degenerate lowest unoccupied orbital and higher energy orbitals.
The two electrons donated by the two Rb atoms per dimer fill this lowest
orbital and a true band insulator is formed.  In this state the magnetic
susceptibility is small, as observed by J\`anossy {\it et al.}.\cite{jan93}
Breaking the dimer apart ({\it e.g.}, by warming) results in a partially
filled band, namely the conducting Rb$_1$C$_{60}$ $fcc$ state.

To conclude, vibrational spectroscopy on Rb$_1$C$_{60}$ has provided strong
evidence for novel structures in what has already been shown to be an
intriguing fullerene compound.  We propose that the quenched phase is
composed of Rb$_2($C$_{60})_2$ dimers and have shown that this structure is
consistent with the IR and Raman data.  Finally, many silent modes of
pure C$_{60}$ , which become activated in either low temperature state of
Rb$_1$C$_{60}$ , are classified as having {\it gerade} or {\it ungerade}
symmetry.

\acknowledgements

We gratefully acknowledge stimulating discussions with P.W. Stephens, L.
Forro, S. Pekker, G. Faigel, G. Oszl\`anyi, M. Pederson, P.B. Allen, and
D.L. Peebles.  We thank D.B. Chrisey for providing the MgO substrates,
and B.V. Shanabrook and D.G. Gammon for allowing us to use their Raman
apparatus.  The work at Stony Brook has been supported by NSF grant DMR
9202528.

\begin{figure}
\caption{Infrared transmission spectra of a Rb$_1$C$_{60}$ sample at several
temperatures in the energy range covering nearly all C$_{60}$
vibrational modes.  The bottom curve is the spectrum of the $fcc$ (high
temperature) phase of the sample.  The middle curve was obtained by
slowly cooling the sample from its $fcc$ state to the polymer state.
The top spectrum
was measured after the sample was rapidly cooled from high temperature
to the new ``quenched" state.  Backgrounds have been subtracted and the
three spectra have been scaled to aid visual comparison.  The center
position of each observed mode is annotated beneath the respective
spectral line.  Small vertical lines above the figure indicate the
positions of the Raman-active $A_g$ and $H_g$ vibrations of pure C$_{60}$.}
\label{IR}
\end{figure}

\begin{figure}
\caption{Raman spectra of a Rb$_1$C$_{60}$ sample in its $fcc$ structure
(bottom panel), its polymer state (middle), and its ``quenched" state (top).
The respective temperatures are indicated on the curves.  The 10
Raman-active modes of pure C$_{60}$ are labeled in the bottom
panel; the frequencies of newly activated modes are labeled for the
quenched and polymer state spectra in the upper two panels.  Inset sketches
represent the dimer and polymer structures of the C$_{60}$ molecules in the
quenched and slow-cooled states of Rb$_1$C$_{60}$ , respectively.}
\label{raman}
\end{figure}

\end{document}